\documentclass[article]{JHEP3}
\usepackage{graphicx}
\usepackage{epsf,colordvi,shadow,pifont}
\usepackage{amsmath,epsfig}
\usepackage{amssymb,amsfonts}
\usepackage{latexsym}
\usepackage{epsfig}
\usepackage[vcentermath,enableskew]{youngtab}
\def\hri#1#2{\href{http://arxiv.org/abs/#1}{[ArXiv:#1]#2}}
\def\hre#1#2{\href{http://arxiv.org/abs/#1/#2}{[ArXiv:#1/#2]}}
\def\hspi#1#2{\href{http://www.slac.stanford.edu/spires/find/hep/www?irn=#1}{#2}}
\relax

\renewcommand{\theequation}{\arabic{section}.\arabic{equation}}

\def\be{\begin{equation}}
\def\ee{\end{equation}}

\newcommand{\bear}{\begin{eqnarray}}
\newcommand{\bea}{\begin{eqnarray}}
\newcommand{\eear}{\end{eqnarray}}
\newcommand{\eea}{\end{eqnarray}}
\def\hre#1#2{\href{http://arxiv.org/abs/#1/#2}{[ArXiv:#1/#2]}}
\def\hspi#1#2{\href{http://www.slac.stanford.edu/spires/find/hep/www?irn=#1}{#2}}
\newbox\pippobox

\renewcommand{\b}[1]{\textbf{#1}}

\def\II{\relax{\rm I\kern-.18em I}}

\def\e{\epsilon}
\def\m{\mu}
\def\n{\nu}
\def\r{\rho}
\def\s{\sigma}
\def\pa{\partial}

\def\sp{\;\;\;,\;\;\;}

\def\a{\alpha}

\def\l{\lambda}
\def\b{\beta}

\title{\huge On novel string theories from 4d gauge theories.\footnote{{\large \it Based on talks given to several institutions as well as workshop and conferences in the last few years.}}}

\author{{\Large Elias Kiritsis$^{a,b}$}\\
~\\
$^a$ \href{http://hep.physics.uoc.gr}{Crete Center for Theoretical Physics},
Departdment of Physics, University of Crete, 71003 Heraklion, Greece\\
~\\
$^b$ \href{http://www.apc.univ-paris7.fr}{APC, Universit\'e Paris 7}, CNRS/IN2P3, CEA/IRFU, Obs. de Paris, Sorbonne Paris
Cit\'e, B\^atiment Condorcet, F-75205, Paris Cedex 13, France (UMR du CNRS 7164).}

\preprint{CCTP-2013-01 }

\abstract{We investigate strings theories as defined from four dimensional gauge theories.
 It is argued that novel (super)string theories exist up to 26 dimensions.
 Some of them may support weakly curved geometries. A proposal is outlined to link their local conformal invariance to the dynamics of the bulk string theory.}


\begin{document}

\section{Introduction}
\setcounter{equation}{0}

Gravity is the first interaction to be successfully modeled in modern science. It also remains todate
probably the least understood. There are two main reasons that  set gravity apart from the other
main known interactions. The first is the fact that the (very successful) classical theory of general relativity
is at best an effective field theory as a quantum field theory (QFT), giving no useful clues on its UV completion.
The second is that there is a major clash between our understanding of QFT and gravity summarized in the cosmological constant
problem, as epitomized by the recent observation of the ``dark energy" filling the universe.

Along with the problems came new ideas on the structure and UV completion of gravity. One input comes from string theories which provide a perturbative
quantum completion of gravity theory, well controlled once energy transfers  remain  much smaller than  the Planck scale.
Another comes from large-N holographic ideas that relate a host of adjoint large-N theories to string theories (containing gravity).

 Closed string theory generically predicts gravity. Fundamental string theories
provide a consistent (perturbative) quantization of gravity.
Despite its successes, string theory, although  well defined at
energies below or at the string
scale, breaks down at energies close to the Planck scale. In particular,
the perturbation theory breaks down due to the strong effective
gravitational coupling. Despite speculations, the nature of the
extreme UV degrees of freedom of
the theory is still obscure.

It was suggested in section 7.8 of \cite{rev} and elaborated further in \cite{talk} that all interactions in
 nature are described by four-dimensional quantum field  theories.
In particular observable gravity is the avatar of a ("hidden"
 in the IR)  large-N gauge theory,
that couples to SM matter via bifundamental fields. Moreover,
a geometrization of this picture suggests the existence of a
(Copernican) chaotic brane universe
where the SM is represented by an appropriate  local stack of
 branes embedded in a higher dimensional universe filled with
 other branes whose relative motion is responsible
to a large degree for the ``local cosmology" observed in our universe, \cite{mirage,rev}.
In a sense, this postulate includes a new level of structure in the cosmos, as well as ``extra dimensions".
Such dimensions,  beyond the four observed ones  are emergent in this picture.

In this postulate, gravitons are composite at high energies and their constituents are ``large-N gluons".
 Today, such a postulate raises few eyebrows due to the Maldacena conjecture
  \cite{mald1,mald2} and subsequent works in the last decade.

Since the early work of 't Hooft \cite{hoof} it was understood that
the low-energy limit of large N-gauge theories is described by some string theory.
The gauge theory versus string theory/gravity
correspondence is a more precise indication that gravity can be realized as an
effective theory of a four-dimensional gauge theory. The inverse
is also true: fundamental string theory in some backgrounds
describes the physics of theories that at low energy are standard
gauge theories. Although bulk-boundary duality is a concept
transcending that of four-dimensional gauge theories, it is most
powerful in the four-dimensional cases.

The lesson of AdS/CFT correspondence is that any
gauge theory has a dual gravity/string theory.
The idea of 't Hooft that gravity must be holographic \cite{holog1}
indicates that a gravity theory must have a dual gauge theory
description.\footnote{We are using the term gauge-theory in a loose sense to include adjoint and bifundamental QFTs.
 There is a distinction between ``color" that stands for a gauge degree of freedom, and ``flavor" that stands for ungauged degrees of freedom.}

 The approach advocated in \cite{rev} has some similarities but also important differences with ideas in
\cite{sundrum1,sundrum2} and \cite{zee}.
The interplay between gauge theory and gravity was suggested by black hole thermodynamics considerations and gave a new view on the thermodynamics of gravity. This view is partly captured by Jacobson's observation that  GR dynamics in many cases is a consequence of the first law of thermodynamics, \cite{jacob} and Verlinde's recent reformulation of gravity as an entropic force, \cite{erik}.

The suggestion, after holography, is that string and gravity theories can be defined by many
QFTs, notably, two-dimensional $\s$-models, four-dimensional
gauge theories, but also three-dimensional Chern-Simons theories and six dimensional
 CFTs.\footnote{In the 3 and 6-dimensional cases the connection to string theory is  more complicated.} 
In this paper we will explore how four-dimensional gauge theories give hints on the existence
 of (super)string theories living in dimensions between 26 and 10 dimensions.\footnote{In \cite{lv} it was argued that in a more general context string theory is the dynamics of sources of quantum field theory. Partial integration of QFT operators generates dynamics of the otherwise classical sources. Similar ideas were independently advocated in \cite{ssl}. }

\section{The many windows to  string theory}

String theories can be constructed in different contexts. The standard approach so far was to construct fundamental string theories in terms of 2-dimensional $\s$-models. We will discuss conventional NSR $\sigma$-models here.\footnote{There have been several other two-dimensional approaches including Green-Schwarz (GS) $\s$-models, \cite{GS} using space-time spinors that are anticommuting world-sheet scalars,  the hybrid formulation, \cite{hyb} containing parts of space-time in the GS formulation and parts in the NSR formulation , as well as the pure spinor formulation, \cite{ps}. The GS formalism is tuned to space-time supersymmetric backgrounds while the others are still being developed and explored.}

In the NSR formulation,   the $\sigma$-model fields, are either bosons (the embedding coordinates of a string) or fermions (fermionic coordinates associated with additional degrees of freedom). Gauge fields in two dimensions are essentially Lagrange multipliers in the continuum (CFT) limit. Therefore, two-dimensional quantum field theories in this class correspond to the NSR formulation in the superstring case.

String backgrounds with non-trivial RR field are not naturally realized in this framework. The RR vertex operators are non-local in terms of the natural variables of the $\sigma$ models (string coordinates), and no technique exists so far to extract physics in this context. Such backgrounds are interesting as they are connected with holographic duals of gauge theories arising from the world volume of D-branes.

Strings and gravity can also emerge from higher dimensional theories. A much studied example is four-dimensional gauge theories, where string coordinates are not the fundamental fields, but strings appear through Wilson loops.
In a sense we can think of strings arising this way as from a quantization of (D)three-branes.

Other theories can give rise to strings. Among them are Chern-Simons-matter theories in three dimensions (related to M2 branes), or the more obscure 6-dimensional theories associated with $M5$ branes.

\subsection{Two-dimensional $\sigma$-models}

The general class of (cutoff) non-linear $\sigma$-models in two dimensions is parametrized by backgrounds associated with massless (graviton, antisymmetric tensor, dilaton) or tachyonic (potential) backgrounds in the bosonic case, while massive backgrounds correspond to irrelevant operators at weak coupling.
In the case of fermionic strings more backgrounds are appearing.

The relevant (dimensionless) coupling of the 2d-theory is the ratio of the string length $\ell_s$, to the volume scale $\ell$ ($volume=\ell^{D}$) of the embedding space in the case of a compact embedding. In non-compact spaces, it is the relevant curvature scale that is replacing the volume.

An independent dimensionless coupling is the string coupling $g_s$, controlling the strength of interactions {\em both at tree level and loop level}.
In a sense the loop expansion is not automatically included in the formulation of the $\s$-model. An additional topological sum must be postulated on varying genus Riemann surfaces weighted by $g_s^{-\chi}$ where $\chi$ is the $Euler$ $number$ of the two-dimensional Riemann surfaces on which the $\sigma$-model is defined.

An important ingredient is conformal invariance, which is necessary if the theory is going to describe a continuum string theory. The relevant conditions for conformal invariance have a simple expansion at weak $\sigma$-model coupling. This expansion is equivalent to an expansion in weakly varying background fields.
For example, the dilaton $\beta$-function reads, \cite{callan}
\be
\beta_{\Phi}=\left(D_b+{1\over 2}D_f\right)-D_{crit}+{3\over 2}{\ell_s^2}\left[4(\nabla\Phi)^2-4\square \Phi-R+{1\over
 12}H^2\right]+{\cal O}(\ell_s^4)
\label{1}\ee
where the critical dimension $D_{crit}$ is 26 for the bosonic string while it is 15 for the fermionic strings. Similarly, $D_b$ are the number of bosonic coordinates while $D_f$ is the number of fermionic coordinates.

At weak coupling, conformal invariance imposes the critical dimension: $\left(D_b+{1\over 2}D_f\right)=D_{crit}$ while curvature corrections are small and the backgrounds are slowly varying.
Subcritical strings, with $\left(D_b+{1\over 2}D_f\right)<D_{crit}$ quickly run to large curvatures and therefore to strong $\sigma$-model coupling. The relevant ``flow" equations (summarized by the two derivative effective action) have AdS-like solutions, as the dilaton potential whose coefficient is
$D_{crit}-\left(D_b+{1\over 2}D_f\right)$ has a positive sign. On the other hand, in the supercritical case  with $\left(D_b+{1\over 2}D_f\right)>D_{crit}$ the equations have deSitter-like solutions.

However, $\sigma$ models do not have an exclusive role in defining string theories. We know at least three other classes of theories that can be a priori used to define string theories: 3-dimensional conformal CS theories, 4-dimensional conformal gauge theories and 6-dimensional exotic conformal fields\footnote{In this case the relevant fundamental object is probably a two-brane that will provide strings upon compactification.}  theories that are poorly and only indirectly  understood.
In the sequel I will focus on four-dimensional gauge theories as an alternative source of string theories.

\subsection{Four-dimensional QFTs}

A weakly coupled four-dimensional QFT can be parametrized in terms of its vector, fermion and scalar degrees of freedom, gauge group and Yukawa and scalar couplings.
UV complete theories are asymptotically conformal or asymptotically free.
They will correspond to critical or subcritical string theories defined via 2d $\s$-models.

To obtain continuum dynamics (and string theories) conformal invariance is necessary as in two dimensions.
Weakly-interacting strings arise when the number of colors is large, \cite{hoof}.

The type of space-times that arise in this context seem to be asymptotically AdS instead of asymptotically flat. Moreover, they also seem to always contain RR backgrounds, a feature that is prohibitive in the standard NSR $\s$-model approach. Indeed it was argued in generality in \cite{disse} that the pseudoscalars, dual to the instanton density, present in any continuum four-dimensional QFT arise in a RR sector
 due to their peculiar $N_c$-dependence, which in string theory translates to the peculiar dilaton dependence of RR-couplings.

In the same vain such theories are superstring theories. They must contain fermionic degrees of freedom, so that they have RR sectors. On the other hand they may not contain space-time fermions, much like the conventional type-0 superstring theories, if there are no fermionic gauge-invariant operators, as it happens for example in pure Yang-Mills.

In this context, the AdS/CFT paradigm suggests that the analogue of the string coupling constant here is the inverse of the number of colors $g_s\sim {1\over N_c}$, while the analogue of the $\s$-model expansion parameter ($\alpha'$) is related to the 't Hooft coupling. For less symmetric gauge theories, this role is played by the many other couplings, including Yukawa's and quartic scalar couplings. Unlike the $\s$-model formulation, the string-loop expansion is tightly interwoven in the definition of the theory as the large-$N_c$ expansion.

There is another main difference with $\s$-models: there seems to be now a finite dimensional space of couplings, at least in perturbation theory. Although this seems to be a major difference, this is probably an illusion. The reason is that (Euclidean) 2d CFTs have always a finite number of relevant or marginal operators. On the other hand in four dimensions there is an extra extension that does resemble somewhat the two-dimensional formulation: making the 4d QFTs couplings depend on the four-dimensional coordinates. There are a few works that have entertained this possibility starting with the standard YM case, \cite{osborn,osborn2}, the N=4 sYM case, \cite{liu,tseytlin},  and the N=1 sYM, \cite{erdmenger}. There is also a recent study of fixed points induced by mildly space-time-dependent couplings, \cite{eva}.

\section{$\beta$ functions for adjoint fields}

Critical theories should correspond to Conformal Theories. In perturbation theory the leading constraints appear at one loop\footnote{This is also the case in the 2d $\s$-model  where the leading contribution to the dilaton $\beta$-function appears at one-loop, \cite{pol}.}.
The one-loop $\beta$-function coefficient for the gauge coupling is given in the appendix in equation (\ref{b1}) and reads
\be
b_1={11\over 3}C_2(G)-{2\over 3}S_2(F)-{1\over 6}S_2(S)+{1\over (4\pi)^2}Y_4(F)
\label{2}\ee
Here $C_2(G)$ is the quadratic Casimir of the adjoint of the group capturing the contribution of vectors, and $S_2(F,S)$ are the Dynkin indices capturing the contributions of charged fermions and bosons (with opposite sign). $Y_4(F)$ is a contribution of Yukawa couplings that is detailed in equation (\ref{b2}) of appendix \ref{apa}.  The associated group representations can be anything, unlike the vectors that transform always in the adjoint. From now on we will assume that the gauge group is SU(N) for concreteness. No major changes are expected for other Lie groups. In particular as we will be interested at theories with a large number of colors, the other relevant groups will be Sp(N) and SO(N) groups. Not surprisingly, they will be associated with unoriented string theories.

A look at the table \ref{table11} in appendix \ref{apa} is enough to convince us that matter representations with Young tableaux with more than two
boxes will completely overwhelm the contributions of the vectors at large $N$. Therefore, only fundamental and bifundamental representations are relevant for near critical theories. The leading contributions to the $\beta$-function will be provided by matter in adjoints, symmetric or antisymmetric representations. We will ignore in the sequel $\Yboxdim8pt\yng(1,1)$ and $\Yboxdim8pt\yng(2)$
    representations for simplicity. They can be taken into account later. They appear again in unoriented string theories and they contribute as half-adjoints to the $\beta$-functions to leading order in $1/N$.

We now consider the presence of $N_F$ Majorana fermions and $N_s$ scalars in the adjoint of SU(N) with no other interactions.
The two-loop $\beta$-function reads
\be
\beta(g)=-{g^3\over (4\pi)^2}\left\{{11\over 3}-{2\over 3}N_F-{N_s\over 6}\right\}N
-{g^5\over (4\pi)^4}\left\{34-16{N_F}-7N_s\right\}{N^2\over 3}+\cdots
\label{3}\ee
We may add Yukawa interactions. We label boson adjoints by $I,J,\cdots$ and fermion adjoints by $i,j,\cdots$.
Then $Y\to Y^I_{ij}{f^a}_{bc}$
and
\be
Y_4(F)={1\over dim(G)}N_f C_2(G)^2dim(G)\sum_{I,i,j}Y^I_{ij}{Y^{I\dagger}}_{ji}=N_FN^2|Y^2|
\label{4}\ee
The $\beta$-function therefore becomes
\be
\beta(g)=-{g^3N\over 6(4\pi)^2}\left\{22-4N_F-N_s\right\}
-{g^5N^2\over 3(4\pi)^4}\left\{34-16{N_F}-7N_s\right\}-{g^3|Y|^2\over (4\pi)^4}N_F N^2\cdots
\label{5}\ee
We also define a 't Hooft-like coupling as $\hat \lambda \equiv {g^2N\over (4\pi)^2}$ to obtain
\be
{\beta_{\hat\l}\over \hat \l^2}=-{1\over 3}\left\{22-4N_F-N_s\right\}
-{2\hat \l\over 3}\left\{34-16{N_F}-7N_s\right\}-2N_F{|Y|^2N\over (4\pi)2}\cdots
\label{6}\ee

The one-loop piece is analogous to the  one-loop piece in the $\s$-model dilaton $\beta$-function in (\ref{1}).
 Like (\ref{1}),  scalars and fermions contribute with the same sign.
Unlike (\ref{1}) which is expressed in a covariant formulation, here we have the analogue of a ``light-cone" formulation. In particular, intuition for the AdS/CFT correspondence indicates that the bosonic adjoint fields will become emerging extra dimensions beyond the four we started with. The vector contribution of 22, is the analogue of the reparametrization-ghost contribution in the $\s$-model\footnote{The vanishing of the one-loop $\beta$ function of a gauge theory with scalar
adjoints and the connection to 26 ``dimensions" was first observed in \cite{fradkin}, where it was pointed out that D-dimensional YM Theory has a vanishing leading correction to its coupling constant for $D=26$. This property was later related in \cite{metsaev} to the special properties of open string theory in D=26 dimensions. The holographic argument linking the one-loop $\beta$-function to 26 dimensions seems different here.}. It would interesting if a ``covariant" formulation exists in this context.

Choosing the number of bosons as $N_s=22-4N_F$ so that the one-loop $\beta$-function vanishes we obtain
\be
{\beta_{\hat\l}\over \hat \l^2}=
{8\hat \l}\left\{10-{N_F}\right\}-2N_F{|Y|^2N\over (4\pi)2}\cdots
\label{7}\ee

For all possible values of $N_F=0,1,2,3,4,5$ allowed by the one-loop condition, a fixed point may be obtained by adjusting the Yukawa couplings.
However, as was argued in \cite{nonc}, using holographic input about the conformal anomaly, it is highly plausible that the only perturbative fixed point with adjoints only is in the family of $N=4$ sYM. All other fixed points are expected at strong coupling, and at large-N we expect to have (super)-gravity duals.

We now proceed to investigate several individual cases.
\begin{itemize}

\item $N_s=22$, $N_F=0$

This theory contains only bosons. The maximal global symmetry here is  O(22), under which the scalars transform in the vector. There are no Yukawa couplings in this case but only quartic scalar coupling.

In a holographic incarnation of conformal invariance it is expected that the O(22) symmetry will be geometrically realized by the emergence of
an $S^{21}$ manifold. Along with an $AdS_5$ realizing conformal symmetry a total of 26 dimensions
are emerging, reminding of the critical dimension of the bosonic string.

The string theory dual to the gauge theory cannot be a purely bosonic string theory, though. As argued extensively in \cite{disse}, in all four dimensional gauge theory duals, the pseudoscalar state (axion) dual to the instanton density, must be a RR state, as defined by the dilaton dependence of its bulk couplings. The presence of RR states indicates a superstring rather than a bosonic string. Of course, the theory in question has only bosonic gauge-invariant operators and therefore the dual superstring theory should not have any space-time fermions (as it happens in type-0 theories).

The two loop $\beta$-function here is not affected by the quartic couplings and has an IR-free sign. However there are AF contributions at three-loops.
Therefore the theory is unstable at weak coupling and this may be a signal of a tachyon in the dual string theory.

\item $N_s=18$, $N_F=1$.
The maximal global symmetry this theory can have is an O(18) for the bosons as well as a U(1) for the fermion. The fermionic U(1) is however anomalous and one is remaining with the O(18) symmetry. If a fixed point is realized at strong coupling without Yukawa couplings then the dual background is expected to be $AdS_5\times S^{17}$ with a total of 22 dimensions.
The presence of Yukawa couplings will introduce an additional 18-dimensional vector and will break $O(18)\to O(17)$. It is expected that this will induce a fibering of an $S^1$ over $S^{16}$  with a radially dependent fiber.

This theory now is expected to have fermionic gauge-invariant operators with simplest ones being spin-1/2 fermions that are bound states of the adjoint fermions and the scalars.

\item $N_s=14$, $N_F=2$.
The maximal global symmetry here is an O(14) symmetry acting on the scalars, as well as a U(2) symmetry acting on the fermions. As usual the U(1) part is anomalous and therefore the surviving fermionic symmetry is SU(2). Were it not for the fermionic symmetry an unbroken scalar symmetry would generate an $AdS_5\times S^{13}$ background with in total 18 dimensions. The presence of Yukawas will embed the SU(2) inside O(14) and will induce a radial fibering of $S^3\times S^{10}$ over $S^{13}$.
This theory also has space-time fermions.

\item $N_s=10$, $N_F=3$.
The maximal global scalar symmetry is here O(10), while we have a non-anomalous SU(3) symmetry of the fermions. The scalar symmetry suggests a maximal $AdS_5\times S^9$ background in 14 dimensions. This symmetry will be deformed by the presence of Yukawa couplings. In this case they will embed $SU(3)\to SU(5)\to O(10)$, which suggest that solution will contain a fibering of
${SU(3)\over SU(2)\times U(1)}\times S^3$ inside $S^9$.

\item $N_s=6$, $N_F=4$. This case contains maximally supersymmetric YM theory, and is probably the only one that exists at weak coupling.
    The maximal scalar symmetry is O(6) and the non-anomalous fermionic one SU(4). The Yukawas identify these two groups that are realized geometrically on an $S^5$, giving a ten-dimensional $AdS_5\times S^5$ background. This provides standard critical superstring theory as well as maximal supersymmetry, although many other ten-dimensional backgrounds provide CFTs with or without supersymmetry.

\item $N_s=2$, $N_F=5$ This is the case with the maximal number of fermions. The maximal fermionic symmetry is SU(5) while the maximal bosonic symmetry is O(2). The minimal realization of the SU(5) symmetry is on the ${SU(5)\over SU(4)\times U(1)}$ eight-dimensional coset. Therefore the expected background is $AdS_5\times {SU(5)\over SU(4)\times U(1)}$ and is thirteen-dimensional. The presence of Yukawa couplings will induce an radial fibering of an $S^1$ in the Grassmannian coset.

\end{itemize}

We observe, that in all cases but the supersymmetric case the expected (super)string theory backgrounds are superficially supercritical (meaning they have $D>10$) although the one loop contribution vanishes. This suggests  the presence of tachyons (associated with IR-free running) at weak coupling.

The  $a-c$ conformal anomaly near the weak coupling fixed point can obtained
from the general expressions
\be
a={2M_s+11M_f+124M_v\over 720(4\pi)^2}\sp c={M_s+3M_f+12 M_v\over 120(4\pi)^2}
\label{8}\ee
where $(M_s,M_f,M_v)$ is the total number of (real)scalars, Majorana fermions and vector degrees of freedom respectively.
Its value for the adjoint theories in question  suggests that only
the supersymmetric D=10 setup with ($N_s=6,N_f=4$) has $a-c=0$ and therefore a conventional
 string dual once the coupling is strong. In all other cases, the knowledge of $a-c$ at a strong coupling fixed point is not known.

We cannot exclude at this point that the strong coupling fixed point, has $a-c\not=0$ which would suggest that the dual string theory does not have
an $\a'\to 0$ limit. Such non-geometric string theories have been argued to exist some time ago, in the NSR framework, \cite{top}, motivated by the existence of CFTs without moduli. A similar phenomenon may be also relevant for string theories with RR backgrounds.

Another issue in the $\s$-model formulation of string theory  is that the presence of world-sheet fermions
must be accompanied with world-sheet supersymmetry, necessary to remove the negative norm degrees of freedom. No way out of this has so far been found, although no proof that it cannot be done otherwise has been given either.
In the gauge theory formulation there seems to be no such constraint in the presence of adjoint fermions.  The reason may be that the gauge theory description is closer to the space-time effective action description where supersymmetry is not necessary.

Another observation is that the different cases of string dimensionalities discussed above
do not seem to be connected via standard RG flows. At least perturbatively decoupling massive
 fields seems to connect these cases to IR-free ones. It is harder to make a statement beyond perturbation theory,
  and it is a very interesting question whether there are non-perturbative RG connections.

\section{Generalized Banks-Zaks fixed points, dressing and flavor-brane condensation\label{BZ}}

The supercritical backgrounds above, with the exception of the 26-dimensional
one can be rendered conformally invariant in gauge perturbation
theory upon addition of fundamental matter.
In the string theory picture this amounts
to introducing flavor branes. \cite{kk}.

Banks and Zaks have argued long-time ago, \cite{bz} that theories where the one-loop
 coefficient of the $\beta$-function can be made hierarchically  smaller and of
  opposite sign from the two-loop coefficient, have a perturbative fixed point.

We consider the gauge theory without  Yukawa or quartic scalar couplings.
Therefore the flavor branes should be thought of as coincident with the 3-branes in the dual string theory picture.
 The general $\beta$-function coefficients are given in (\ref{b1}) and (\ref{b2}).
We will look for situations where  the one-loop coefficient of the $\beta$-function
can be made hierarchically  smaller and of opposite sign from the two-loop coefficient, \cite{bz}.
We define the integer $D$ as
\be
{11\over 3}-{2\over 3}N_F-{N_s\over 6}={D\over 6}\sp 0\leq D\leq {22}
\label{b19}\ee
We include in the gauge theory $n_F$ fundamental fermions and $n_s$ fundamental scalars.
We choose their number  so that the integer $\e$ is positive and O(1).
\be
b_1={D\over 6}N-{2\over 3}n_F-{n_s\over 6}={\ell\over 6} >0 \sp \ell\ll N
\label{b20}\ee
This can be easily achieved when $N$ is large.

Then
\be
b_2=\left[{34-16{N_F}-7N_s\over 3}-{13\over 12}D\right]N^2
-{n_s\over 4N}(N^2-3) +{\cal O}\left({\ell\over N}\right)=
\label{b21}\ee
$$
=-\left[{50+4N_F+5N_s\over 4}N^2+{n_s\over 4N}(N^2-3)\right] ++{\cal O}
\left({\ell\over N}\right)<0
$$
for $N>1$.

For ${\ell\over N}\ll 1$ there is a Bank-Zaks-like fixed point at
\be
{\l_*\over (4\pi)^2}={g^2_*N\over (4\pi)^2}\simeq {2\over 3}{N\e\over (50+4N_F+5N_s)N^2+{n_s\over N}(N^2-3)}
\label{b22}\ee
The case  associated with the maximal emergent dimension is obtained for $N_F=0,N_s=21$, where $D=1$
and $\ell={N}-4n_F-n_s$
This can arranged  for example by taking $n_F=0$ and $n_s=N-1$, so that $\ell=1$ and
\be
{\l_*\over (4\pi)^2}\simeq {2\over 3}{1\over 55N+(N-1){N^2-3\over N^2}}\simeq
 {1\over 84N}+{\cal O}(N^{-2})
\label{b23}\ee
This is a reliable fixed point in perturbation theory at large $N$.

These arguments suggest that for each choice of adjoints so that the one-loop
$\beta$-function is negative, there should be an associated conformal window,
in analogy with that of generalized QCD. It therefore provides a finite set of
 families of conformal theories labeled by ($N_F,N_s$).
 As discussed in  appendix \ref{BZYM}, by varying the masses of the fundamentals, one can interpolate between
the BZ-like fixed points and the AF theories parameterized by  ($N_F,N_s$) with exponentially small scales.
At large $N_c$ this is generic and such theories may still be considered continuum string theories, albeit with a graviton mass $m_g\sim e^{-{1\over g_s}}$
that is exponentially small in perturbation theory. This is rather district from graviton masses that scale as $m_g\sim g_s$ associated to string theories interacting with multi-trace interactions \cite{aharony}.

We may calculate the values of the central charges (controlling the conformal anomaly)
for such fixed points from (\ref{8}) to obtain
\be
2880(4\pi)^2(a-c)=9(6-N_s)(N_c^2-1)-(9n_s-7\ell)N_c-7
\label{b2}\ee
Choosing $n_s\simeq (6-N)s)N_c+{\cal O}(N_c^0)$ we obtain $a-c\simeq {\cal O}(N_c^0)$ in the weakly coupled theory.
 If, like in N=1 sQCD the $a-c$ in the IR CFT is independent of flavor, such theories
  would be candidates for a weakly curved string theory dual.

The contribution of fundamentals in rendering the gauge theory conformally invariant is
 reminiscent of the role of the Liouville mode in subcritical string theory but is distinct in its details.

\section{Identifying the string}

There are important differences between 2d $\s$-models and 4d gauge theories when it comes to identify the string degrees of freedom.
In the first case, the 2d quantum fields are directly identified with the coordinates of the string, $X^{M}$ (and their fermionic partners $\psi^{M}$ when present).
In the 4d case the relation is less direct. The natural (gauge invariant) string degree of freedom in a gauge theory is the Wilson loop
\be
W(C)\equiv Tr[{\cal P}e^{i\int_C A_{\mu}\partial_s x^{\mu}ds}]
\label{s1}\ee
However, this is a (closed) loop in 4 dimensions.  A generalization of the Wilson loop in the general gauge theory is necessary in order to capture the
extra (emergent) dimensions expected in the dual string description.
 In the simplest case with 22 adjoint bosons only, $\phi^I$, the generalization of the Wilson
is following studies in the AdS/CFT correspondence,
\be
W({\cal C})\equiv Tr[{\cal P}e^{i\int_C (A_{\mu}(x(s))\partial_s x^{\mu}+\sum_{I=1}^{22}\phi^I(x(s))\theta^I(s))ds}]
\label{s2}\ee
Now the path ${\cal C}$ is a path in a 26-dimensional space spanned by the 4 standard space-time dimensions $x^{\mu}$ as well as 22 extra dimensions $x^I$ with
$\partial_s x^I=\theta^I$.
As the associated Wilson loop in the AdS/CFT correspondence is pointing to a
 Wilson loop in ten-dimensional super-YM, similarly here the generalized
 Wilson loop in (\ref{s2}) is suggestive of a Wilson loop in 26-dimensional pure YM theory,
as anticipated for different reasons in \cite{fradkin}.

The Wilson loop can then be converted into an analogue of the first quantized action for string propagation by considering the amplitude for a Wilson loop ${\cal C}_t$, at time $t$ to evolve to ${\cal C}_{t'}$ at time  $t'$,
\be
\langle W({\cal C}_t)|  W({\cal C}_{t'})\rangle_c=\int {\cal D}A_{\mu}{\cal D}\phi^I~W({\cal C}_t)W({\cal C}_{t'})~e^{-S_{\rm gauge ~theory}}
\label{s3}\ee

The case where adjoint (Majorana) fermions $\psi^I_{\alpha}$ (I is a multiplicity index, $\a$ is a spinor index) are present is a bit trickier.
The intuition coming from the AdS/CFT correspondence suggests that the fermion number symmetry associated to the fermions should be realized geometrically
and therefore from the 4-dimensional point of view the fermionic number symmetry should be a Kaluza-Klein symmetry.

Therefore, a Wilson loop in order to wander in this extra ``fermionic dimension", it must have non-trivial fermion number.
The simplest such Wilson loop (of length (= charge) $m$), that wanders in a single fermionic dimension (that is associated for example with $\psi^1$ is therefore
\be
W_{f}(m)\equiv Tr\left(\psi^1_{\a_1}\psi^1_{\a_2}\psi^1_{\a_{3}}\cdots \psi^1_{\a_{m}}\right)
\label{s4}\ee
Each of the adjoint fermion operators above transforms in the $\left({1\over 2},0\right)$ representation of the Lorentz group.
There are several ways of contracting spinor indices that  make $W_f$ a 4d scalar.
The operators in (\ref{s4}) are of the same type as the long-string operators
that have been discussed in the context of the standard AdS/CFT correspondence.
In any gauge theory a common set of operators are generated by powers of the YM field strength as
\be
W_{F}(m)\equiv Tr\left(F_{\m_1\nu_1}F_{\m_2\n_2}\cdots F_{\m_{m}\n_{m}}\right)
\label{s5}\ee
A basis of self-dual (1,0) and anti-self-dual (0,1) components can be chosen as, \cite{ferretti}
\be
F^{\m\n}=\phantom{\left[\right.}{{\bar\s} _{\m\n}}\phantom{\left.\right]} ^{{\dot\a\dot\b}} ~{\bar f}_{\dot\a\dot\b}+{\s_{\m\n}}^{\a\b}~f_{\a\b}
\label{s6}\ee
with
\be
\s_{\m\n}={i\over 4}\s^2(\s_{\m}\bar \s_{\n}-\s_{\n}\bar \s_{\m})\sp \bar\s_{\m\n}=-{i\over 4}(\bar\s_{\m}\s_{\n}-\bar \s_{\n}\s_{\m})\s^2
\label{s7}\ee
Unlike (\ref{s4}) the individual ``string bits" in (\ref{s5}) have SU(2) spin 1 instead of ${1\over 2}$.

As shown in \cite{ferretti}, the mixing of the operators (\ref{s5}) in pure YM theory is described by an integral spin chain where two types of spins are involved: and SU(2) spin 1/2 where ($f,\bar f$) transform as a doublet and a SU(2) spin 1,
reflecting the spin content of the (1,0) and (0,1) representations
of ($f,\bar f$). Intuition from the $N=4$ superYM and the AdS-CFT correspondence suggests that the full dynamics of the long ``string-bit" operators in the continuum limit gives the non-linear string $\s$-model.

We have given earlier a general argument that the string theories dual to 4d gauge theories
 are ``superstrings" with fermionic world-sheet degrees of freedom.
From the discussion above we do expect world-sheet fermionic degrees of freedom
 as they arise in general in the continuum limit of spin chains.
This is a hint that the world-sheet description of the emergent strings theory
 involved fermions (superstrings) or their duals (spins).
Indeed, in an NSR formulation, the vertex operators
for RR states are spin-fields expected to emerge from a continuum limit of spin chains.

\section{Discussion}

We have made an analogy between NSR $\sigma$-models in two dimensions
and gauge theories in four dimensions, in terms of their ability to define continuum string theories.
Based on this analogy we have argued for the existence of novel (super)string
theories in discrete dimensions lower than 26.

Most of the requisite fixed points we are after are at strong gauge coupling and are therefore unknown.
We have managed to construct however non-trivial fixed points in all
dimensions but 26, using the BZ idea (implying the condensation of flavor branes
in the dual string theory picture).

At weak gauge coupling the dual string theories are expected to be strongly curved,
 but we have given evidence that there may be examples also of weakly curved duals.
Proceeding further and unraveling such theories is a non-trivial challenge.

An interesting question is whether there is an expansion in gauge theories that is the analogue of the $\sigma$ model expansion
in (\ref{1}) at the two-loop order and beyond. To do this one would have to turn on the natural space-time dependent couplings of the
marginal (massless) operators of YM theory. The generic such operators are classified by the Young tableau of O(4) and are quadratic in the field strength.
They are $Tr[F_{\m\n}F_{\r\s}]$ minus all traces ($\Yboxdim8pt\yng(2,2)$) , as well as  the ``traces"
 associated with the traceless stress tensor $T_{\m\n}=Tr[F_{\m\r}{F_{\n}}^{\rho}]-{1\over 4}\eta_{\m\n}Tr[F^2]$, ($\Yboxdim8pt\yng(2)$), the action
 $Tr[F^2]$ (singlet) and the topological density (a pseudo-singlet), $Tr[F\wedge F]$, ($\Yboxdim8pt\yng(1,1,1,1)$).
The $\Yboxdim8pt\yng(2,2)$ operator does not correspond to a
``massless" state as we understand from the AdS/CFT correspondence,
but for all the others we can turn on space-time dependent couplings that are the space-time metric $g_{\m\n}$, the gauge coupling constant (alias the dilaton, $e^{\phi}$) and the $\theta$-angle, (alias the axion).
We will therefore consider an action of the form
\be
S_{YM}=-{1\over 4}\int d^4x \sqrt{g}Tr\left[e^{-\phi(x)}~g^{\m\r}g^{\n\s}F_{\m\n}F_{\r\s}-i{\theta(x)\over 8\pi^2} F\wedge F\right]
 \label{d1}\ee
with $\phi,\theta,g_{\m\n}$ slowly varying functions of $x$.
The scale and Weyl invariance, in the presence of non-trivial space-time dependent couplings (or sources)
 has been partly analyzed in the recent past, \cite{osborn}-\cite{tseytlin}.

 The variation of the effective functional $W$ of a 4d QFT  in the presence of sources $g_i$ ( space-time dependent couplings) has the following
 general form, \cite{osborn}
\be
\delta_{\s}W(g_i)=\int d^4x\sqrt{g}\left[\s\sum_i\beta_i {\delta W\over \delta g_i}+\s L_C+\partial_{\m}\s Z^{\m}\right]
\label{d2}\ee
where $\s$ is the infinitesimal parameter of a conformal transformation. The local terms depending on the metric and couplings are
\be
L_{C}=\beta_a W^2+\b_b GB+{1\over 9}\beta_c R^2+{1\over 3}\chi^1_i\partial_{\m}R+{1\over 6}\chi_{ij}^2 \pa_{\m}g_i\pa^{\m}g_jR
+{1\over 2}\chi_{ij}^3 \pa_{\m}g_i\pa_{\n}g_j G^{\m\n}+
\label{d3}\ee
$$
+{1\over 2}\chi_{ij}^4\square g_i\square g_j+{1\over 2}\chi^5_{ijk}\pa_{\m}g_i\pa^{\m}g_j\square g_j+{1\over 4}\chi^{6}_{ijkl}
\pa_{\m}g_i\pa^{\m}g_j
\pa_{\m}g_k\pa^{\m}g_l
$$
with
\be
W^2=R_{\m\n;\r\s}R^{\m\n;\r\s}-2R_{\m\n}R^{\m\n}+{1\over 3}R^2\sp GB=R_{\m\n;\r\s}R^{\m\n;\r\s}-4R_{\m\n}R^{\m\n}+R^2
\label{d4}\ee
\be
G_{\m\n}=R_{\m\n}-{1\over 2}Rg_{\m\n}
\label{d5}\ee
The terms appearing in $Z^{\m}$ can be removed by redefinitions of the couplings as well as of the functional $W$, and we will assume that they are zero from now on. The various terms appearing in (\ref{d3}) are constrained by the abelian nature of the Weyl transformations, \cite{osborn}.

For the pure YM theory in (\ref{d1}) a one-loop calculation gives, \cite{osborn2}, after redefinitions
\be
L_C={1\over 4Im\tau^2}\left(D^2 \tau D^2\bar \tau-2G^{\m\n}\nabla_{\m} \tau\nabla_{\n}\bar \tau-{R\over 3}\nabla^{\m} \tau\nabla_{\m}\bar \tau\right)
+{( \nabla^{\m} \tau\nabla_{\m}\bar \tau)^2 \over 16Im\tau^4}+
\label{d6}\ee
$$+{\nabla^{\m} \tau\nabla^{\n} \tau\nabla_{\m}\bar \tau\nabla_{\n}\bar \tau-( \nabla^{\m} \tau\nabla_{\m}\bar \tau)^2\over 48Im\tau^4}
$$
where
\be
\tau={\theta\over 2\pi}+{4\pi i e^{-\phi}}\sp D^2\tau=\square \tau+{i\over Im\tau}\nabla^{\m}\tau\nabla_{\m}\tau
\sp D^2\bar \tau=\square \bar \tau-{i\over Im\tau}\nabla^{\m}\bar \tau\nabla_{\m}\bar\tau
\label{d7}\ee
At this order the SL(2,R)-invariance is manifest.

In the two-dimensional formulation of string theory, such equations are equivalent to the space-time equations for the background field.
The situation in four dimensions seems different. The conformal invariance here as well is expected to relate the ``world-volume dynamics" (of the gauge theory) to the bulk string dynamics. The Weyl anomaly equations above do not contain all the emergent string dimensions and the associated generalized background fields. They do capture however the divergent terms of the on-shell bulk
 effective action in the higher dimensions, via the usual AdS-CFT correspondence.
  Therefore, formulae like (\ref{d3}) or (\ref{d6}) give constraints on the bulk effective action.
 It is unclear at the moment whether they can determine with precision the bulk effective action. We can however imagine the following line of thought: 

\begin{itemize}

\item The general procedure of holographic renormalization in the presence of several bulk fields, \cite{papa}
maps the bulk action (with two derivatives)
to the generalized conformal anomaly. Moreover in this formulation, the quadratic and quartic divergences
 are manifest\footnote{These are absent from (\ref{d2}), (\ref{d3}) as such calculations are done in dimensional regulation that is not capable of capturing power divergences.}.

\item The arguments above suggest that the UV divergences of the QFT, as captured in the generalized conformal anomaly calculation above, map directly to IR divergences from the bulk string action.

\item The calculation of the finite parts (namely to leading order the effective potential for scalars),
\cite{niarchos},  indicates that it contains apart from regularity of the holographic solutions the same information as that necessary to compute the divergent parts.

\item This suggests that in the context of holography, the generalized Weyl anomaly in
 four dimensions including all divergences may be able, as in two dimensions
to generate the on-shell bulk (string) dynamics, once an ansatz is made for the higher-dimensional effective string theory.
If this turns out to be correct, it will put the four dimensional case on a par with the standard two dimensional formulation.

\end{itemize}

There are some obvious lines of research in view of the above.

\begin{enumerate}

\item Use non-trivial 4d CFTs to determine the associated string dynamics. Such CFTs must be non-free in order to provide non-trivial examples of string dynamics. Weakly coupled CFTs of the BZ type may be good candidates for this.

\item Work out in detail the algorithm presented above, linking the generalized conformal anomaly of the CFT to the IR divergences of the bulk string theory therefore determining the effective action.

\end{enumerate} 

\vskip 1.2cm
\centerline{\bf Acknowledgments}
\vskip 1.2cm
\noindent
I would like to acknowledge useful discussions with K. Intriligator, E. Rabinovici and A. Schwimmer.
Parts of this work have been reported earlier in \cite{talk}, in seminars at Perimeter, IPMU and CERN, as well as at the Planck conference, in June 2010.

 This work was in part supported by grants  PERG07-GA-2010-268246, PIEF-GA-2011-300984, and the EU program
``Thales" ESF/NSRF 2007-2013. It has also been
co-financed by the European Union (European Social Fund, ESF) and Greek national
funds through the Operational Program ``Education and Lifelong Learning" of the National
Strategic Reference Framework (NSRF) under ``Funding of proposals that have received a
positive evaluation in the 3rd and 4th Call of ERC Grant Schemes".

\vskip 1.2cm

\vskip 1.2cm
\centerline{\bf Note added}
\vskip 1.2cm
\noindent

After this work was posted on the archive, A. Tseytlin has informed us of past work in some of the issues advocated here.
In particular, earlier observations of a connection of the the one-loop $\beta$-function to 26 dimensions and the open bosonic string, \cite{fradkin,metsaev}
as well as work on the conformal invariance of 4d gauge theories in the presence of space-time couplings, \cite{osborn}-\cite{tseytlin}.
Finally the prescient paper of Lovelace, \cite{love} has constructed a string theory (at tree level) starting from a a conformal field theory in any dimension. Moreover, he argued how the UV divergences of the CFT map to the IR divergences of the spacetime physics that control the dynamics of light modes.
This has been works out only however for 2-dimensional CFTs.

 \newpage
\appendix

 \renewcommand{\theequation}{\thesection.\arabic{equation}}
\addcontentsline{toc}{section}{Appendices}
\section*{APPENDIX}

\section{Gauge theory $\beta$-functions\label{apa}}

The $\beta$ function for the gauge coupling constant is to two loops, \cite{jones,mv}
$$
\beta(g)\equiv \mu{dg\over d\mu}=-{g^3\over (4\pi)^2}\left\{
{11\over 3}C_2(G)-{2\over 3}S_2(F)-{1\over 6}S_2(S)+{1\over (4\pi)^2}Y_4(F)\right\}-
$$
\be
-{g^5\over (4\pi)^4}\left\{{34\over 3}(C_2(G))^2-\left[2C_2(F)+{10\over 3}C_2(G)\right]S_2(F)-
\left[2C_2(S)+{1\over 3}C_2(G)\right]S_2(S)\right\}+\cdots
\label{b1}\ee
where $F$ is the fermion representation (two-component spinors), $S$ the (real) scalar representation,
and $Y_4(F)$ is the contribution of possible Yukawa couplings
\be
Y_4(F)={1\over dim(G)}Tr[C_2(F)Y^aY^{a\dagger}]
\label{b2}\ee
 In particular for the Yukawa coupling $Y^a_{ij}$,  $a$ labels boson representation indices while $i,j$ fermion ones.

We use hermitian representation matrices for the color group $G$ satisfying
\be
[T^a_{R},T^b_{R}]=i {f^{ab}}_{c} T^c_{R}
\label{b3}\ee
in the representation $R$.
The quadratic Casimir $C_2(R)$ of the representation $R$ is defined as
\be
(T^a_RT^a_R)_{ij}=C_2(R)\delta_{ij}
\label{b4}\ee
It should be distinguished from the Dynkin index $S_2(R)$ defined
\be
Tr[T^a_RT^b_R]=S_2(R)\delta^{ab}
\label{b5}\ee
They are related by
\be
d(G)S_2(R)=d(R)C_2(R)
\label{b6}\ee
where $d(R)$ is the dimension of $R$ and $d(G)$ is the dimension of the adjoint.

We have the following relations
\be
f^{acd}{f^{b}}_{cd}=C_2(G) \delta^{ab}
\label{b8}\ee
which defines the normalization convention and
\be
-if^{abc}T^a_RT^b_RT^c_R={1\over 2}C_2(G)C_2(R)\sp T^a_RT^b_RT^a_RT^b_R=\left[C_2(R)-{1\over 2}C_2(G)\right]C_2(R)
\label{b9}\ee
and
\be
-if^{acd}Tr[T^b_RT^c_RT^d_R]={1\over 2}C_2(G)S_2(R)\delta^{ab}
\sp
Tr[T^a_RT^c_RT^b_RT^c_R]=\left[C_2(R)-{1\over 2}C_2(G)\right]S_2(R)\delta^{ab}
\label{b10}\ee
The values for these indices for low lying SU(N) representations can be found in table \ref{table11}.

\begin{table}[!t]
\begin{center}
\begin{tabular}{|c|c|c|c|} \hline
Representation & dimension & Dynkin Index $S_2$ & Casimir $C_2$\\
\hline \hline

$\Yboxdim8pt\yng(1)$ & $N$ & ${1\over 2}$ & ${N^2-1\over 2N}$\\

$\Yboxdim8pt\yng(2)$   &  ${N(N+1)\over 2}$ & ${N+2\over 2}$ & ${(N-1)(N+2)\over N}$\\

$\Yboxdim8pt\yng(1,1)$   &  ${N(N-1)\over 2}$ & ${N-2\over 2}$ &${(N+1)(N-2)\over N}$\\

Adjoint &  $N^2-1$& $N$&$N$\\

$\Yboxdim8pt\yng(3)$& ${N(N+1)(N+2)\over 6}$& ${(N+2)(N+3)\over 4}$&${3(N-1)(N+3)\over 2N}$ \\

$\Yboxdim8pt\yng(2,1)$& ${N(N^2-1)\over 3}$& ${N^2-3\over 2}$&${3(N^2-3)\over 2N}$\\

$\Yboxdim8pt\yng(1,1,1)$& ${N(N-1)(N-2)\over 6}$& ${(N-2)(N-3)\over 4}$&${3(N+1)(N-3)\over 2N}$  \\

\hline
\end{tabular}
\label{table11}
\caption{Relevant data for some low-lying SU(N) representations}
\end{center}
\end{table}

\subsection{$\beta$ functions for fundamentals and adjoints}

We will now consider an ``orientable" spectrum containing apart from the gauge field in the adjoint of SU(N),   $N_F$ Majorana fermions and $N_s$ (real) scalars in the adjoint of SU(N),   as well as $n_F$
 fundamental fermions+antifermions and $n_s$ fundamental scalars+antiscalars
with no other interactions.
The $\beta$-function (\ref{b1}) now reads
\be
\beta(g)=-{g^3\over (4\pi)^2}b_1-{g^5\over (4\pi)^4}b_2+{\cal O}(g^7)
\label{b11}\ee
with
\be
b_1=\left\{{11\over 3}-{2\over 3}N_F-{N_s\over 6}\right\}N-{2\over 3}n_F-{n_s\over 6}
\label{b12}\ee
\be
b_2=\left\{34-16{N_F}-7N_s\right\}{N^2\over 3}-n_F\left({N^2-1\over N}+{10\over 3}N\right)
-n_s\left({N^2-1\over N}+{N\over 3}\right)
\label{b13}\ee

We now add Yukawa interactions for the adjoints. We label boson adjoints by $I,J,\cdots$ and fermion adjoints by $i,j,\cdots$.
Then ${Y^{Ia}}_{bi,cj}\to Y^I_{ij}{f^a}_{bc}$
and from (\ref{b2})
\be
Y_4(F)={1\over dim(G)} C_2(G)^2dim(G)\sum_{I,i,j}Y^I_{ij}{Y^{I\dagger}}_{ji}=N^2|Y^2|
\label{b14}\ee
The adjoint Yukawa contribution to the $\beta$-function therefore becomes
\be
\Delta\beta(g)=-{g^3|Y|^2\over (4\pi)^4}N^2\cdots
\label{b15}\ee
It can cancel positive contributions coming from the ${\cal O}(g^5)$ terms.

We now move to the $\beta$-functions for Yukawa couplings. At one-loop
\be
(4\pi)^2\beta^a_1={1\over 2}Y^{\dagger}_2(F)Y^a+{1\over 2}Y^aY_2(F)+2Y^bY^{a\dagger}Y^b+Y^bTr[Y^{b\dagger}Y^a]-
3g^2\{C_2(F),Y^a\}
\label{b16}\ee
where
\be
Y_2(F)= Y^{\dagger a}Y^a\sp
Y_2 (S) = Tr[Y^{\dagger a}Y^b]= Y_2(S)\delta^{ab}
\label{b16a}\ee
For the adjoint couplings we obtain
\be
Y_2(F)_{ia,jb}=-C_2(G)\delta^{ab}M_{ij}\sp M_{ij}=\sum_{I=1}^{N_s}\sum_{k=1}^{N_F}Y^{I,\dagger}_{ik}Y^I_{kj}=\sum_{I=1}^{N_s}\sum_{k=1}^{N_F}(Y^{I}_{ki})^*Y^I_{kj}
\ee
\be
 Y_2(S)^{Ia,Jb}=-C_2(G)\delta^{ab}N^{IJ}\sp N^{IJ}=\sum_{i,j=1}^{N_F}Y^{I,\dagger}_{ij}Y^J_{ji}=\sum_{i,j=1}^{N_F}(Y^{I}_{ji})^*Y^J_{ji}
\ee
and we also have
\be
{f^{ca}}_{d}{f_{be}}^d{f^b}_{fc}={c_2(G)\over 2}{f^a}_{ef}
\ee
and substituting in (\ref{b16}) we have
\be
(4\pi)^2(\beta_1)^I_{ij}={N\over 2}\left[\sum_{j=1}^{N_F}(M_{ik}Y^I_{kj}+Y^{I}_{ik}M_{kj})+2\sum_{J=1}^{N_s}\sum_{k,l=1}^{N_F}Y^J_{ik}(Y^I_{lk})^*Y^J_{lj}+2\sum_{J=1}^{N_s}
Y^J_{ij}N^{JI}\right]-6g^2N Y^I_{ij}
\label{b17}\ee

There can be also quartic couplings
\be
V_4\equiv {\l_{abcd}\over 4!}\Phi^a\Phi^b\Phi^c\Phi^d=
\ee
$$
{\lambda_1\over 4!} Tr[\Phi^I\Phi^J\Phi^I\Phi^J]+{\lambda_2\over 4!} Tr[\Phi^I\Phi^I\Phi^J\Phi^J]+
{\lambda_3\over 4!} (Tr[\Phi^I\Phi^I])^2+{\lambda_4\over 4!} Tr[\Phi^I\Phi^J]Tr[\Phi^I\Phi^J]
$$
The case $\l_2=-\l_1$, $\l_3=\l_4=0$ corresponds to the potential of ${\cal N}=4$ SYM.
We will ignore the double trace terms in the sequel.
The one-loop $\beta$-function is
\be
(4\pi)^2\beta_{abcd}=\Lambda^2_{abcd}-4H_{abcd}+\Lambda^Y_{abcd}-3g^2\Lambda^S_{abcd}+3g^4A_{abcd}
\ee
with
\be
A_{abcd}={1\over 8}\sum_{perm}\{T^A,T^B\}_{ab}\{T^A,T^B\}_{cd}\sp H_{abcd}=
{1\over 4}\sum_{perm}Tr[Y^aY^{b\dagger}Y^cY^{d\dagger}]
\ee
\be
\Lambda^2_{abcd}={1\over 8}\sum_{perm}\l_{abef}\l_{efcd}\sp \Lambda^Y_{abcd}=\sum_{k=a,b,c,d}Y_2(k)\lambda_{abcd}
\sp \Lambda^S_{abcd}=\sum_{k=a,b,c,d}C_2(k)\lambda_{abcd}
\ee
where the permutation sum is over the 4! permutations of the indices $a,b,c,d$.

Here as there are no fermions the $H_{abcd}$ and $\Lambda^Y_{abcd}$ terms as absent.
We parametrize
\be
\lambda_{Ia,Jb,Kc,Ld}=\left[\l_1\delta_{IJ}\delta_{KL}+\l_2\delta_{IK}\delta_{JL}\right]Tr[T^aT^bT^cT^d]_{sym}
\ee
with $T^a_{bc}={f^a}_{bc}$.

\section{IR-Modified Banks-Zaks theories\label{BZYM}}

We will investigate in this appendix the modifications to the BZ story of section \ref{BZ} once the fundamentals are massive. To consider a naturally RG stable setup we will assume that the fundamental fields are fermions.

We will first start by writing again the BZ $\beta$-function and the associated flow. This assumes that the fundamentals are massless. We use as variable, $\hat\lambda={g^2N\over (4\pi)^2}$
\be
\beta_{BZ}(\hat\lambda)=-\e\hat\lambda^2+b_*\hat\lambda^3+\cdots
\sp
\left({\e\over b_*\hat\lambda(\mu)}-1\right)e^{\e\over b_*\hat\lambda(\mu)}=\left({\mu\over \Lambda}\right)^{\e^2\over b_*}
\label{b24}\ee
In the BZ region, $\e={\cal O}\left({1\over N}\right)\ll 1$, $b_*={\cal O}(1)$.
We have
\be
\hat\lambda(\mu\to\infty)={1\over \e\log {\mu\over \Lambda}}+\cdots\sp \hat\lambda(\mu\to 0)={\e\over b_*}-{e^{-1}\e\over b_*}\left({\mu\over \Lambda}\right)^{\e^2\over b_*}+\cdots
\label{b25}\ee
The scale $\Lambda$ is controlling the flow to the BZ fixed point, and the associated gauge theory. At $\mu\ll \Lambda$ the theory is in the scaling region of the IR fixed point. At  $\mu\gg \Lambda$ the theory in the AF region controlled by $\Lambda$.

We now consider the fundamentals having a common mass $m$. Qualitatively, for $\mu\gg m$ the flavors are effectively massless, and the flow is as above. Below $m$ however the flavors decouple and the theory is asymptotically free with a one loop $\beta$ function $b_0={\cal O}(1)$.
In this regime the flow is controlled by a YM-like $\beta$-function. We will assume a sharp transition
 at $\mu=m$ to simplify the flow. For $\mu\ll m$ we obtain
\be
\beta_{YM}(\hat\lambda)=-b_0\hat\lambda^2+\cdots\sp {1\over \hat\lambda(\m)}=
{1\over \hat\lambda(m)}+b_0\log{\mu\over m}
\label{b26}\ee
where $\hat\lambda (m)$ is evaluated from the BZ flow
\be
\left({\e\over b_*\hat\lambda(m)}-1\right)e^{\e\over b_*\hat\lambda(m)}=\left({m\over \Lambda}\right)^{\e^2\over b_*}
\label{b27}\ee

In this case the theory in the ultimate IR is AF, and the coupling is driven to infinity. We can calculate the effective IR scale associated with the AF running of the coupling as
\be
\Lambda_{IR}=m~e^{-{1\over b_0\hat\lambda (m)}}
\label{b28}\ee
We will now estimate the range of $\Lambda_{IR}$ as a function of ${m\over \Lambda}$.

For ${m\over \Lambda}\ll 1$, we use (\ref{b25}) to obtain
\be
\Lambda_{IR}\sim m~e^{-{b_*\over b_0\e}}\sp  {m\over \Lambda}\ll 1
\ee
while in the opposite regime
\be
{\Lambda_{IR}\over \Lambda}\sim \left({m\over \Lambda}\right)^{1-{\e\over b_0}} \sp  {m\over \Lambda}\gg 1
\ee

We observe that in the BZ regime $\e\ll 1$, for low masses, one obtains a YM scale hierarchically smaller than $m$.
From this point of view when $m\ll \Lambda$, the quarks are ``heavy" from the point of view of IR YM theory as they satisfy $m\gg \Lambda_{IR}$. In this theory the spectrum of low-lying  glueballs is expected to have  characteristic masses of order $\Lambda_{IR}$ while the mesons will have masses of order $m$.

The theory is AF in the perturbative regime, then flows very close to the perturbative BZ fixed point up to energy $m$, where the coupling starts to go away from the BZ fixed point theory and to flow to the YM IR. When $m\ll \Lambda$ the energy region $\Lambda\gg E\gg m$ is a ``walking" region for the gauge theory. The theory is AF and perturbative between the $m$ and $\Lambda_{IR}$ scales.
The three scales $\Lambda$, $m$ and $\Lambda_{IR}$ can be naturally chosen to be exponentially away one from the other.

Once we leave the BZ regime, and while we are in the conformal window,
we expect the picture above to persist, although now the two scales $\Lambda_{IR}$ and
 $m$ will be of the same order of magnitude.


\end{document}